\documentclass[twocolumn,abs,prb,longbibliography]{revtex4-1}

\usepackage{graphicx}
\usepackage{dcolumn}
\usepackage{float}
\usepackage{amsmath}
\usepackage{bm}

\newcommand{\comment}[1]{}

\begin{document}


\title{Signatures of merging Dirac points in optics and transport}

\author{J. P. Carbotte$^{1,2}$}
\author{E. J. Nicol$^{3}$}
\email{enicol@uoguelph.ca}
\affiliation{$^1$Department of Physics and Astronomy, McMaster
University, Hamilton, Ontario L8S 4M1, Canada}
\affiliation{$^2$The Canadian Institute for Advanced Research, Toronto, ON M5G 1Z8, Canada}
\affiliation{$^3$Department of Physics, University of Guelph,
Guelph, Ontario N1G 2W1, Canada} 
\date{\today}

\begin{abstract}{
    We consider the optical and transport properties in a model two-dimensional Hamiltonian which describes the merging of two Dirac points.
    At low energy, in the presence of an energy gap parameter $\Delta$, there are two distinct Dirac points with linear dispersion, these are connected by a saddle point at higher energy.
    As $\Delta$ goes to zero, the two Dirac points merge and the resulting dispersion exhibits semi-Dirac behaviour which is quadratic in the $x$-direction (``nonrelativistic'') and linear the $y$-direction (``relativistic'').
    In the clean limit for each direction ($x,y$) the contribution of the intraband and interband optical transitions are both given by universal functions of photon energy $\Omega$ and chemical potential $\mu$ normalized to the energy gap. We provide analytic formulas for both small and large $\Omega/2\Delta$ and $\mu/\Delta$ limits. These define, respectively, Dirac and semi-Dirac-like regions. For $\Omega/2\Delta$ and $\mu/\Delta$ of order one, there are deviations from these asymptotic behaviors. Considering optics and also transport, such as dc conductivity, thermal conductivity and the Lorenz number, such deviations provide signatures of the evolution from the Dirac to the semi-Dirac regime as the gap $\Delta$ is varied. 
}
\end{abstract}

\maketitle

\section{Introduction}

Following the isolation of graphene in 2004, there has been intense interest
in its electronic properties.\cite{CastroNeto:2009} This was soon followed with the discovery of topological insulators\cite{Hasan:2010,Qi:2011} and more recently three-dimensional Dirac and Weyl semimetals\cite{Armitage:2018}, materials which exhibit many novel exotic properties. The field of topological materials represents a very active new and exciting frontier in condensed matter physics.\cite{Yan:2017,Wehling:2014,Chiu:2016} Optical studies of electronic systems have provided a wealth of  important and detailed information on charge dynamics including the high $T_c$ superconducting cuprates\cite{Basov:2005,Carbotte:2011}, graphene\cite{Li:2008,Carbotte:2010}, topological insulators\cite{Schafgans:2012}, Dirac and Weyl materials\cite{Chen:2015,Xu:2016,Neubauer:2016}, to give a few examples. The experimental work has also been informed and supported by many theoretical studies including works on graphene\cite{Carbotte:2010}, on Dirac and Weyl semimetals\cite{Tabert:2016a,Tabert:2016b,Carbotte:2016} and multi-Weyl\cite{Ahn:2017}.

The possibility of the merging of Dirac points in two dimensional crystals has received much attention.\cite{Wunsch:2008,Hasegawa:2006} A universal Hamiltonian to describe this situation was introduced by Montambaux {\it et al.}\cite{Montambaux:2009a,Montambaux:2009b} which was subsequently used to describe many properties associated with this Hamiltonian. These include the calculation of the electronic density of states and specific heat\cite{Montambaux:2009b}, Bloch-Zener oscillations\cite{Lim:2012}, interband tunneling\cite{Fuchs:2012}, the role of winding numbers\cite{Gail:2012}, Friedel oscillations\cite{Dutreix:2013}, the corresponding Hofstadter spectrum\cite{Delplace:2010}, screening and plasmons\cite{Pyatkovskiy:2016}, interplay between topology and disorder\cite{Sriluckshmy:2018}, Hall viscosity and relation to Berry curvature\cite{Pena:2019}. We also note an experimental realization in optical lattices\cite{Tarruell:2012} and another in microwave cavities\cite{Bellec:2013}.

In this paper, we consider both optical and transport properties. Recently Adroguer {\it et al.}\cite{Adroguer:2016} have considered in the diffusive limit some aspects of the transport and optics using both a Boltzmann and a diagrammatic approach with particular emphasis on anisotropy in the residual scattering that has its origin in the semi-Dirac nature of the model. In one direction, the band structure is ``relativistic'' (energy is linear in momentum), while in the other it is ``nonrelativistic'' (energy is quadratic in momentum). Ziegler {\it et al.}\cite{Ziegler:2017} have considered the optical conductivity within a tight-binding model which includes a high energy van Hove singularity. They also discuss the diffusion regime. In another recent work, Mawrie and Muralidharan\cite{Mawrie:2019} calculate the optical conductivity numerically from a Kubo formula and present results for one particular set of parameters characterizing the Hamiltonian which depends on an energy gap $\Delta$, an effective mass $m$ describing motion in the $x$-direction and a velocity $v$ for the relativistic propagation in the $y$-direction. A specific value of the residual scattering rate is also used.

Here, we will use the continuum model of Ref.~\onlinecite{Montambaux:2009b} and focus on the clean limit. We will show that the conductivity can be reduced to universal forms valid for any value of $m$, $\Delta$, and $v$. The forms are different for the longitudinal conductivity in the $x$ and $y$ directions and for inter- and intraband transitions. In each case, they are reduced to a single integral over angle with an integrand analogous to that which enters elliptic integrals.

The necessary formalism is presented in section II, where our continuum Hamiltonian is specified and the Kubo formula for the optics is given. The clean limit is introduced in section III along with formulas for inter- and intraband contributions to the real part of the dynamic conductivity, reduced to universal forms involving a single integral applicable to any value of $\Delta$ which simply scales the energy. Analytic results are provided in some simplifying limits. In section IV, we present our numerical results for the interband part of the conductivity as a function of photon energy normalised to $2\Delta$.
We compare our results with our analytic results for the limit of $\Omega\to 0$, which reduces to a pure Dirac behavior, while in the large $\Omega$ limit semi-Dirac\cite{Carbotte:2019} applies. 
In section~V, we turn our attention to the variations in transport coefficients as the magnitude of the gap is varied and the transition to semi-Dirac progresses. We treat the dc electrical conductivity, the thermal conductivity, and the Lorenz number, and present
results both as a function of doping (chemical potential) at zero temperature and for variation with temperature at zero doping.
In section VI, we provide our final summary and conclusions.

\section{Theoretical Formalism}

\begin{figure}
\includegraphics[width=0.9\linewidth]{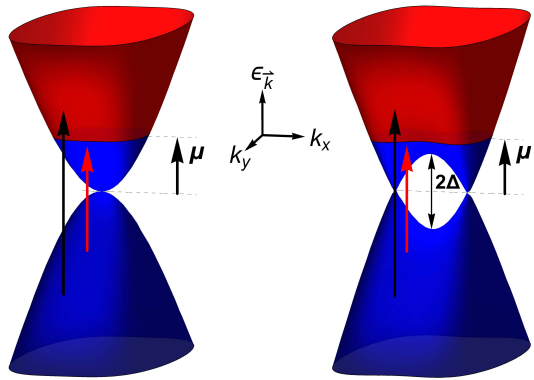}
\caption{A schematic of the dispersion curves $E_{\bm k}$ vs ${\bm k}$ associated with the Hamiltonian of Eq.~(\ref{eq:Hmatrix}). The left side is for $\Delta=0$ (semi-Dirac case) and the right for $\Delta\ne 0$. The upper bands above zero for the chemical potential are the conduction bands and the lower ones, the valence bands. Here, the blue regions are for occupied states and the red, unoccupied.
The black arrows show allowed optical transitions and the red, blocked transitions. 
}\label{fig1}
\end{figure}

This work is based on the two-dimensional continuum Hamiltonian used to describe the merging of two Dirac points into one. It has the form\cite{Montambaux:2009a,Montambaux:2009b}
\begin{equation}
\hat H=\left(\begin{array}{cc}
0 & \displaystyle\frac{\hbar^2k_x^2}{2m}-\Delta-i\hbar vk_y\\
\displaystyle\frac{\hbar^2k_x^2}{2m}-\Delta +i\hbar vk_y & 0
\end{array}\right) ,
\label{eq:Hmatrix}
\end{equation}
which reduces to the semi-Dirac form in the limit of zero gap ($\Delta=0$). In this limit, the dispersion curves are quadratic in the $x$-direction and linear in the $y$-direction. The energy dispersion $E_{\bm k}$ has the form
$E_{\bm k}=\pm\sqrt{({\hbar^2k_x^2\over 2m}-\Delta)^2+\hbar^2v^2k^2_y}$, 
with $v$ the ``relativistic''  velocity and $m$ the effective mass associated with the $x$-direction. Fig.~\ref{fig1} provides a schematic representation of this energy dispersion for finite $\Delta$ and $\Delta=0$, the latter being the usual semi-Dirac case. Note that semi-Dirac dispersions have been theoretically demonstrated in a variety of systems: TiO$_2$/V$_2$O$_3$ nanostructures\cite{Pardo:2009,Banerjee:2009,Pardo:2010,Banerjee:2012}, photonic materials\cite{Wu:2014}, and hexagonal lattices under a magnetic field\cite{Dietl:2008}.

The conductivity can be computed from a Kubo formula which depends only on the electron spectral density matrix $\hat A({\bm k},\omega)$ associated with the chosen Hamiltonian in Eq.~(\ref{eq:Hmatrix}). The conductivity is typically discussed in terms of its longitudinal  versus transverse response relative to the orientation of the applied electric field, for example $\sigma_{xx}$ and $\sigma_{xy}$, respectively. In general, the component of the real part of the conductivity $\sigma_{ij}(T,\Omega)$ with temperature $T$ and photon energy $\Omega$ takes the form
\begin{eqnarray}
\sigma_{ij}(T,\Omega)=& \displaystyle N_f\hbar\frac{ e^2}{2\Omega}\int^{+\infty}_{-\infty}{d\omega\over 2\pi} [f(\omega) - f(\omega+\Omega)]\nonumber\\
&\displaystyle\times\int \frac{d^2{k}}{4\pi^2} {\rm Tr}[\hat v_i \hat A({\bm k},\omega) \hat v_j \hat A({\bm k},\omega+\Omega)]
\label{eq:condformula}
\end{eqnarray}
with $N_f$ a degeneracy factor which here we take to be 1, while for graphene it is 4 (associated with valley and spin degeneracy), $e$ is electron charge and 
$f(\omega)$ is the Fermi-Dirac function $f(\omega)=1/(1+{\rm exp}[(\omega-\mu)/k_BT])$, where $\mu$ is the chemical potential, and $k_B$ is the Boltzmann constant. The $\hat v_i$ are velocity matrices defined as 
\begin{equation}
\hat v_i ={1\over\hbar}{\partial \hat H\over\partial k_i}
\label{eq:v}
\end{equation}
and the electron spectral functions are related to the Green's function $G(\bm k,z)$ by
\begin{equation}
\hat A(\bm k,\omega)=-2{\rm Im}\hat G(\bm k,z=\omega+i0^+),
\end{equation}
with
\begin{equation}
\hat G^{-1}(\bm k,z)=z\hat I-\hat H,
\end{equation}
where $\hat I$ is the identity matrix.

Detailed examples exist in the literature of manipulating these equations to arrive at a final form for the conductivity, for instance, the case of AA-stacked bilayer graphene,\cite{Tabert:2012} silicene with a tunable gap,\cite{Stille:2012}, and a semi-Dirac $\alpha$-T$_3$ model\cite{Bryenton:2018}. Thus, following a similar sequence of steps, and 
after considerable, but standard, algebra, the $xx$ component of the conductivity takes the form
\begin{eqnarray}
\sigma_{xx}(T,\Omega)&= \displaystyle N_f\frac{ e^2}{h}\frac{4}{\Omega\sqrt{2mv^2}}\int^{+\infty}_{-\infty}d\omega [f(\omega) - f(\omega+\Omega)]\nonumber\\
&\times\displaystyle\int_0^\infty d\epsilon\,  {\epsilon^{3/2}}\int_0^{\pi/2}d\phi\sqrt{\cos\phi}[B_1I_{\rm intra}+B_2I_{\rm inter}],\nonumber\\
& \label{eq:sigmaxx}
\end{eqnarray}
where 
\begin{eqnarray}
I_{\rm intra}&=& A(\omega-E)A(\omega+\Omega-E)
\nonumber\\
&&\qquad +A(\omega+E)A(\omega+\Omega+E),\label{eq:Iintra}\\
I_{\rm inter}&=& A(\omega-E)A(\omega+\Omega+E)
\nonumber\\
&&\qquad +A(\omega+E)A(\omega+\Omega-E),
\label{eq:Iinter}
\end{eqnarray}
with $A(x)= (1/\pi)(\Gamma/[x^2+\Gamma^2])$, which introduces a constant impurity parameter $\Gamma$,
and
\begin{eqnarray}
B_1=& \displaystyle 1-{\epsilon^2\over E^2}\sin^2\phi,\\
B_2=& \displaystyle {\epsilon^2\over E^2}\sin^2\phi.
\end{eqnarray}
With regard to the form of the spectral function $A(x)$, in many-body theory, impurity scattering enters through a complex self-energy $\Sigma$ which shifts the electronic energy by ${\rm Re} \Sigma$ and gives a quasiparticle lifetime from ${\rm Im} \Sigma$. Consequently, $A(x)=|{\rm Im} \Sigma|/[(x-{\rm Re} \Sigma)^2+({\rm Im} \Sigma)^2]/\pi$. We use the simplest model of impurity scattering which is phenomenological, where ${\rm Im} \Sigma=\Gamma$ and ${\rm Re} \Sigma=0$. It is used extensively  throughout the literature when discussing optical and transport properties. 

The first term in the square brackets of Eq.~(\ref{eq:sigmaxx}) describes the intraband (Drude) processes and the second term, the interband optical transitions. In obtaining Eq.~(\ref{eq:sigmaxx}), we have chosen to transform coordinates from $(k_x,k_y)$ to new variables $(\epsilon,\phi)$ with 
$\epsilon^2\equiv(\hbar^2k_x^2/2m)^2+(\hbar vk_y)^2$ and $\phi=\tan^{-1}[(\hbar v k_y)/(h^2k^2_x/2m)]$. For the $yy$ component of the conductivity, we obtain
\begin{eqnarray}
\sigma_{yy}(T,\Omega)&= \displaystyle N_f\frac{ e^2}{h}\frac{\sqrt{2mv^2}}{\Omega}\int^{+\infty}_{-\infty}d\omega [f(\omega) - f(\omega+\Omega)]\nonumber\\
&\times\displaystyle\int_0^\infty d\epsilon  {\sqrt{\epsilon}}\int_0^{\pi/2}d\phi{1\over\sqrt{\cos\phi}}[B_2I_{\rm intra}+B_1I_{\rm inter}].\nonumber\\
& \label{eq:sigmayy}
\end{eqnarray}
Note that 
\begin{eqnarray}
E&\equiv&\displaystyle\sqrt{\biggl({\hbar^2k_x^2\over 2m}-\Delta\biggr)^2+\hbar^2v^2k^2_y}\nonumber\\
&\equiv&\sqrt{\Delta^2+\epsilon^2-2\Delta\epsilon\cos\phi}.
\end{eqnarray} 

\begin{figure}
\includegraphics[width=0.9\linewidth]{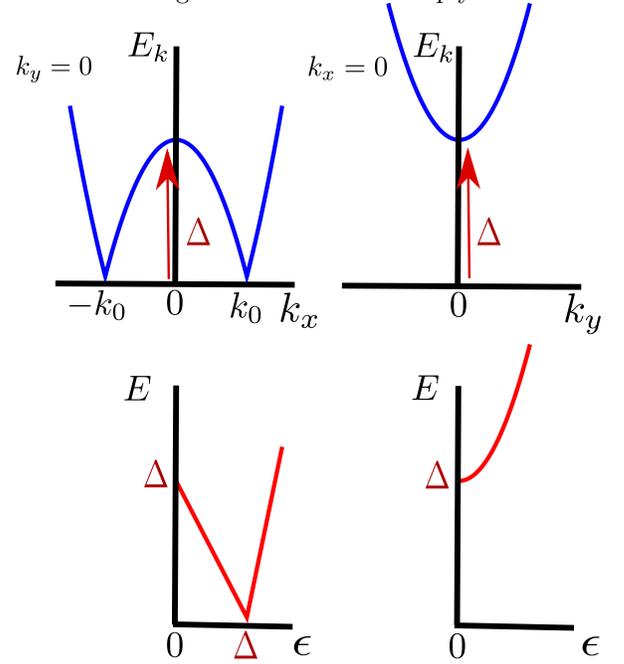}
\caption{
A schematic of the dispersion curve cross sections for positive energy. The top frames are for the variables $(k_x,k_y)$ with $(k_x,0)$ shown on the left and $(0,k_y)$, on the right. The bottom frames are for comparison and show the transformed variables $(\epsilon,\phi)$ with $\phi=0$ on the left and $\phi=\pi/2$ on the right.
}\label{fig2}
\end{figure}

In Fig.~\ref{fig2}, we show cross sections of the dispersion energy for the conduction band. On the left side of the figure, we show the variation of $E$ as a function of $k_x$ for $k_y=0$ in the upper frame and on the right hand side, we show it as a function of $k_y$ for $k_x=0$. We note the qualitative differences between the two frames. The right hand frame shows a pure gapped Dirac behavior which for $\Delta=0$ goes into a linear dispersion $\hbar vk_y$ at large $k_y$. The left frame is very different. It shows a van Hove singularity at $k_x=0$ and two zeroes at $k_x=\pm k_0=\pm \sqrt{2m\Delta/\hbar^2}$. The velocity out of these nodes is $v_x=\sqrt{2\Delta/m}$ which in general differs from $v_y=v$ and $v_x=\hbar k_x/m$.
In the lower frame of this same figure, we show the energy $E$ but as a function of the transformed coordinates $\epsilon$ and $\phi$ for two values of $\phi$, namely, 0 and $\pi/2$. The right frame is for $\phi=\pi/2$ and here $E$ shows a gapped Dirac behavior $E=\sqrt{\Delta^2+\epsilon^2}$ while in the left frame for $\phi=0$, $E=|\Delta-\epsilon|$ and the dispersion curves are linear in $\epsilon$. 

\section{Clean limit for interband conductivity}

In the clean limit ($\Gamma\to 0$), the spectral functions in Eqs.~(\ref{eq:Iintra}) and (\ref{eq:Iinter}) reduce to Dirac delta functions. We begin with the interband optical transitions. These functions reduce to $[\delta(E+{\Omega\over 2})+\delta(E-{\Omega\over 2})]\delta(\omega+{\Omega\over 2})/2$ which allows the $\omega$ integration to be done simply and as $E$ and $\Omega$ are both positive, the first delta function in the square bracket contributes zero. At $T=0$ and taking $\mu\ge 0$, we obtain
\begin{widetext}
\begin{eqnarray}
\sigma_{xx}^{\rm inter}(T=0,\Omega)&=& \displaystyle N_f\frac{ e^2}{h}\frac{2}{\Omega\sqrt{2mv^2}}\theta(\Omega-2\mu)
\int_0^\infty d\epsilon\,  {\epsilon^{3/2}}\int_0^{\pi/2}d\phi\sqrt{\cos\phi}
{\epsilon^2\over E^2}\sin^2\phi\,\delta(E-{\Omega\over 2})
 \label{eq:sigmaxxinter}\\
\sigma_{yy}^{\rm inter}(T=0,\Omega)&=& \displaystyle N_f\frac{ e^2}{h}\frac{\sqrt{2mv^2}}{2\Omega}\theta(\Omega-2\mu)\int_0^\infty d\epsilon  {\sqrt{\epsilon}}\int_0^{\pi/2}d\phi{1\over\sqrt{\cos\phi}}\biggl[1-{\epsilon^2\over E^2}\sin^2\phi\biggr]\,\delta(E-{\Omega\over 2}).\label{eq:sigmayyinter}
\end{eqnarray}
For the intraband processes, the product of the two spectral density factors in  Eq.~(\ref{eq:Iintra}) reduce to $\delta(\Omega)
[\delta(\omega-E)+\delta(\omega+E)]$ 
and we obtain in the limit of $T\to 0$
\begin{eqnarray}
\sigma_{xx}^{\rm intra}(T,\Omega)&=& \displaystyle N_f\frac{ e^2}{h}\frac{4}{\sqrt{2mv^2}}\delta(\Omega)
\int^{\infty}_{-\infty}-{\partial f(\omega)\over\partial\omega}d\omega
\int_0^\infty d\epsilon\,  {\epsilon^{3/2}}\int_0^{\pi/2}d\phi\sqrt{\cos\phi}
\biggl[1-{\epsilon^2\over E^2}\sin^2\phi\biggr]\,[\delta(\omega-E)+\delta(\omega+E)]\nonumber\\
&& \label{eq:sigmaxxintra}\\
\sigma_{yy}^{\rm intra}(T,\Omega)&=& \displaystyle N_f\frac{ e^2}{h}\sqrt{2mv^2}\delta(\Omega)\int^{\infty}_{-\infty}-{\partial f(\omega)\over\partial\omega}d\omega
\int_0^\infty d\epsilon  {\sqrt{\epsilon}}\int_0^{\pi/2}d\phi{1\over\sqrt{\cos\phi}}{\epsilon^2\over E^2}\sin^2\phi\,[\delta(\omega-E)+\delta(\omega+E)].
\label{eq:sigmayyintra}
\end{eqnarray}
\end{widetext}
At zero temperature $-\partial f(\omega)/\partial\omega$ becomes a delta function $\delta(\omega-\mu)$ and for positive chemical potential ($\mu\ge 0$), the second delta function in the last square bracket in each of Eq.~(\ref{eq:sigmaxxintra}) and (\ref{eq:sigmayyintra}) contributes zero to the integral over $\omega$. This integral can be done by replacing $\omega$ by $\mu$ leading to the single delta function $\delta(E-\mu)$. To do the remaining integral over $\epsilon$ in Eqs.~(\ref{eq:sigmaxxinter})-(\ref{eq:sigmayyintra}), it is convenient to change variable from $\epsilon$ to $E$. The quadratic equation relating $\epsilon$ to $E$ has two solutions
\begin{equation}
\epsilon_\pm=\Delta\cos\phi\pm\sqrt{E^2-\Delta^2\sin^2\phi}.
\end{equation}
Because $\epsilon$ is real, we need the condition $E\ge \Delta\sin\phi$ and for $\epsilon<\Delta\cos\phi$, we have to take $\epsilon_-$ while for $\epsilon>\Delta\cos\phi$, $\epsilon_+$ is needed. After some careful algebra, we arrive at
\begin{widetext}
\begin{eqnarray}
\sigma_{xx}^{\rm inter}(\Omega)&=& \displaystyle N_f\frac{ e^2}{h}4\sqrt{{\Delta\over 2mv^2}}\theta(\Omega-2\mu)\biggl\{{1\over 4\bar\Omega^2}
\int_0^{\pi/2}d\phi\sqrt{\cos\phi}\sin^2\phi
{\theta(\bar\Omega-\sin\phi)\over\sqrt{\bar\Omega^2-\sin^2\phi}}[\bar\epsilon^{7/2}_-\theta(2\Delta-\Omega)+\bar\epsilon^{7/2}_+]\biggr\}
\label{eq:sxxinter}\\
\sigma_{yy}^{\rm inter}(\Omega)&=& \displaystyle N_f\frac{ e^2}{h}\sqrt{mv^2\over 2\Delta}\theta(\Omega-2\mu)\biggl\{{1\over 2\bar\Omega^2}\int_0^{\pi/2}d\phi{1\over\sqrt{\cos\phi}}
{\theta(\bar\Omega-\sin\phi)\over\sqrt{\bar\Omega^2-\sin^2\phi}}\nonumber\\
&\times&[\sqrt{\bar\epsilon_-}
(\bar\Omega^2-\bar\epsilon^2_-\sin^2\phi)\theta(2\Delta-\Omega)
+\sqrt{\bar\epsilon_+}
(\bar\Omega^2-\bar\epsilon^2_+\sin^2\phi)]\biggr\}
\label{eq:syyinter}
\end{eqnarray}
with $\bar\epsilon_\pm=\cos\phi\pm\sqrt{(\Omega/2\Delta)^2-\sin^2\phi}$ and $\bar\Omega=\Omega/2\Delta$.
For the intraband optical transitions, we obtain
\begin{eqnarray}
\sigma_{xx}^{\rm intra}(\Omega)&=& \displaystyle N_f\frac{ e^2}{h}4\Delta\sqrt{{\Delta\over 2mv^2}}\delta(\Omega)
\biggl\{\int_0^{\pi/2}d\phi\sqrt{\cos\phi}
{\theta(\bar\mu-\sin\phi)\over\bar\mu\sqrt{\bar\mu^2-\sin^2\phi}}[\bar\epsilon^{3/2}_-(\bar\mu^2-\bar\epsilon^2_-\sin^2\phi)\theta(\Delta-\mu)+\bar\epsilon^{3/2}_+(\bar\mu^2-\bar\epsilon^2_+\sin^2\phi)]\biggr\}\nonumber\\
\label{eq:sxxintra}\\
\sigma_{yy}^{\rm intra}(\Omega)&=& \displaystyle N_f\frac{ e^2}{h}\sqrt{2mv^2\Delta}\delta(\Omega)\biggl\{
\int_0^{\pi/2}d\phi{1\over\sqrt{\cos\phi}}
{\theta(\bar\mu-\sin\phi)\over\bar\mu\sqrt{\bar\mu^2-\sin^2\phi}}\sin^2\phi[\bar\epsilon^{5/2}_-\theta(\Delta-\mu)+\bar\epsilon^{5/2}_+]\biggr\}
\label{eq:syyintra}
\end{eqnarray}
with $\bar\epsilon_\pm=\cos\phi\pm\sqrt{\bar\mu^2-\sin^2\phi}$ with $\bar\mu=\mu/\Delta$.
\end{widetext}

It is convenient to denote the four universal functions defined in the curly brackets in Eqs.(\ref{eq:sxxinter})-(\ref{eq:syyintra}) by 
${\cal F}_{xx}(\bar\Omega)$,
${\cal F}_{yy}(\bar\Omega)$,
${\cal G}_{xx}(\bar\mu)$ and
${\cal G}_{yy}(\bar\mu)$, respectively. 
These functions have simple analytic forms for $\bar\Omega\to 0$ and $\bar\Omega\to\infty$.
We find:
\begin{eqnarray}\displaystyle
{\cal F}_{xx}(\bar\Omega\to 0) &={\pi\over 8},\quad {\cal F}_{xx}(\bar\Omega\to \infty) &={C_-^{xx}\over 8\sqrt{2}}\sqrt{\Omega\over\Delta},\label{eq:Fxx}\\ 
{\cal F}_{yy}(\bar\Omega\to 0)&={\pi\over 4}, \quad {\cal F}_{yy}(\bar\Omega\to \infty) &={C_+^{yy}\over 2\sqrt{2}}\sqrt{\Delta\over\Omega},\label{eq:Fyy} \\
{\cal G}_{xx}(\bar\mu\to 0)&={\pi\over 2}\bar\mu, \quad {\cal G}_{xx}(\bar\mu\to \infty) &={C_+^{xx}\over 2}\bar\mu^{3/2}, \label{eq:Gxx}\\
{\cal G}_{yy}(\bar\mu\to 0)&={\pi\over 2} \bar\mu,\quad {\cal G}_{yy}(\bar\mu\to \infty) &={C_-^{yy}\over 2}\bar\mu^{1/2}. \label{eq:Gyy}
\end{eqnarray}
To convert ${\cal F}_{xx}$, ${\cal F}_{yy}$, ${\cal G}_{xx}$ and ${\cal G}_{yy}$ 
to conductivity, we need only multiply by the factors that appear in front of the curly brackets in Eqs.~(\ref{eq:sxxinter})-(\ref{eq:syyintra}).
These asymptotic forms have proved to be very useful in the discussion of our numerical results based on Eqs.~(\ref{eq:sxxinter})-(\ref{eq:syyintra}).
Here,
\begin{eqnarray}
 C^{xx}_\pm &=C_1^{xx}\pm C_2^{xx},\\
C_\pm^{yy} &=C_1^{yy}\pm C_2^{yy},
\label{eq:Cpm}
\end{eqnarray}
where
\begin{eqnarray}\displaystyle
 C_1^{xx} &=&\displaystyle\int^{\pi/2}_0\sqrt{\cos\phi} d\phi = {1\over G},\\
 C_2^{xx} &=&\displaystyle\int^{\pi/2}_0\sqrt{\cos\phi}\cos 2\phi d\phi = {1\over 5G},\\
 C_1^{yy} &=&\displaystyle\int^{\pi/2}_0\displaystyle{1\over\sqrt{\cos\phi}} d\phi = \pi G,\\
 C_2^{yy} &=&\displaystyle\int^{\pi/2}_0\displaystyle{\cos 2\phi\over\sqrt{\cos\phi}} d\phi =-{\pi G\over 3},
\label{eq:Cs}
\end{eqnarray}
with the Gauss constant $G\approx0.8346$.
When the asymptotic forms from Eqs.~(\ref{eq:Fxx})-(\ref{eq:Gyy}) are used in the equations for the conductivity, they provide an important link to Dirac and semi-Dirac materials. For example, in the limit of $\Omega\to 0$, $\sigma_{xx}^{\rm inter}(\Omega\to 0)=N_f(\pi e^2/4h)\sqrt{2\Delta/mv^2}\theta(\Omega-2\mu)$
and $\sigma_{yy}^{\rm inter}(\Omega\to 0)=N_f(\pi e^2/4h)\sqrt{mv^2/2\Delta}\theta(\Omega-2\mu)$. Both
$\sigma_{xx}^{\rm inter}$
and $\sigma_{yy}^{\rm inter}$ are constant in this regime with the first proportional to $\sqrt{\Delta}$ which increases with $\Delta$, and the second goes like $1/\sqrt{\Delta}$ which decreases as $\Delta$ increases. The ratio of these two variations goes as $2\Delta/mv^2$ and the square root of the product is equal to $\pi e^2/4h$ ($N_f=1$) which is independent of the material parameters characterizing our Hamiltonian in Eq.~(\ref{eq:Hmatrix}), namely, the mass $m$, gap $\Delta$ and velocity $v$.
It
behaves like graphene for which the value of the constant universal background conductivity is $\pi e^2/2h$, but this is for a degeneracy factor of 4. Here we have taken $N_f=1$ but as can be seen in Fig.~\ref{fig2} left top frame, here we are dealing with two Dirac nodes and so we agree with graphene.

In the limit of large $\Omega\to\infty$, we recover the semi-Dirac behavior described by our Hamiltonian with $\Delta=0$ in this case: 
$\sigma_{xx}^{\rm inter}(\Omega\to\infty)=N_f(e^2C_-^{xx}/4h)\sqrt{\Omega/mv^2}\theta(\Omega-2\mu)$ and $\sigma_{yy}^{\rm inter}(\Omega\to\infty)=N_f(e^2C_+^{yy}/4h)\sqrt{mv^2/\Omega}\theta(\Omega-2\mu)$, where we have used Eqs.~(\ref{eq:sxxinter}), (\ref{eq:syyinter}), (\ref{eq:Fxx}) and (\ref{eq:Fyy}). The gap has dropped out of each of these quantities and they properly reduce to the known results.\cite{Carbotte:2019} In semi-Dirac $\sigma_{xx}^{\rm inter}$ goes like $\sqrt{\Omega}$  and $\sigma_{yy}^{\rm inter}$ goes like $1/\sqrt{\Omega}$. As we have noted before for the limit of $\Omega\to 0$, here for $\Omega\to\infty$ the material parameters $m, v, \Delta$ also drop out of the square root of the product $\sqrt{\sigma_{xx}^{\rm inter}\sigma_{yy}^{\rm inter}}$ 
which is equal to $N_f{e^2\sqrt{C_-^{xx}C_+^{yy}}/4h}$. 
This differs from the Dirac case by the numerical factor of ${\sqrt{C_-^{xx}C_+^{yy}}/\pi}$.

\begin{figure}
\includegraphics[width=0.9\linewidth]{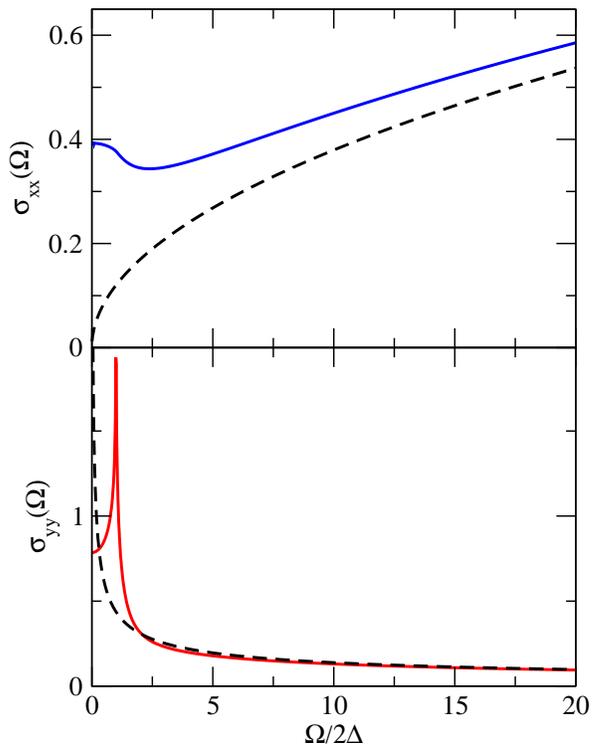}
\caption{The interband contribution to the dynamic conductivity $\sigma^{\rm inter}(\Omega)$ as a function of photon energy normalized to $2\Delta$. The top frame is for $\sigma_{xx}^{\rm inter}(\Omega)$ 
(blue curve) while the bottom frame applies to $\sigma_{yy}^{\rm inter}(\Omega)$ (red curve), in units of $N_f(e^2/h)2\sqrt{2\Delta/mv^2}$ 
and $N_f(e^2/h)\sqrt{mv^2/2\Delta}$, respectively.  Here, the chemical potential has been taken to be zero (charge neutrality). The black dashed curves are for comparison and are the results in the semi-Dirac case [$\Delta=0$ in the Hamiltonian of Eq.~(\ref{eq:Hmatrix})]. These are given by $0.12\sqrt{\Omega/2\Delta}$ and $0.437\sqrt{2\Delta/\Omega}$, using the $\bar\Omega\to\infty$ limits of Eq.~(\ref{eq:Fxx}) and (\ref{eq:Fyy}), respectively. 
}\label{fig3}
\end{figure}

\section{Results for inter- and intraband conductivity}

In Fig.~\ref{fig3}, we show our results for the interband conductivity 
$\sigma_{xx}^{\rm inter}(\Omega)$ and $\sigma_{yy}^{\rm inter}(\Omega)$
as a function of $\Omega/2\Delta$ in the top frame (solid blue curve) and  bottom frame (solid red curve), respectively. These results are based on Eqs.~(\ref{eq:sxxinter})  and (\ref{eq:syyinter}), respectively.
Similar behavior has also been found by Mawrie and Muralidharan\cite{Mawrie:2019}, where they have considered one specific set of values for the parameters versus our generic scaled curves shown here.
While $\sigma_{xx}$ has a change from convex to concave at $\Omega=2\Delta$, $\sigma_{yy}$ displays a van Hove singularity. This can be traced to the behavior of the electronic dispersion curves. In the top frame of Fig.~\ref{fig2}, we note a peak at $k=0$ in the left diagram and a minimum in the right. The density of states $N(\omega)$ itself also has a van Hove singularity which, in this case, is at $\omega=\Delta$. An analytic formula for $N(\omega)$ in terms of elliptical integrals is given in Ref.~\onlinecite{Montambaux:2009b}. Here we use
\begin{eqnarray}
N(\omega)&=&{N_f\over \pi^2\hbar^2}\sqrt{2m\over v^2}\omega\int_0^{\pi/2}{d\phi\over\sqrt{\cos\phi}}{\theta(\omega-\Delta\sin\phi)\over\sqrt{\omega^2-\Delta^2\sin^2\phi}}\nonumber\\
&&\qquad\times[\sqrt{\epsilon_+}+\sqrt{\epsilon_-}\theta(\Delta-\omega)],
\label{eq:dos}
\end{eqnarray}
where $\epsilon_\pm=\Delta\cos\phi\pm\sqrt{\omega^2-\Delta^2\sin^2\phi}$.
For $\omega\to 0$
\begin{equation}
N(\omega)={N_f\over \pi\hbar^2}\sqrt{2m\Delta\over v^2}{\omega\over\Delta},
\label{eq:dosdirac}
\end{equation}
which is the density of states for a Dirac material. For $\omega\to\infty$,
\begin{equation}
N(\omega)={N_f\over \pi^2\hbar^2}\sqrt{2m\Delta\over v^2}C_1^{yy}\sqrt{\omega\over\Delta},
\label{eq:dossemidirac}
\end{equation}
which is the result for a semi-Dirac material.

\begin{figure}
\includegraphics[width=0.9\linewidth]{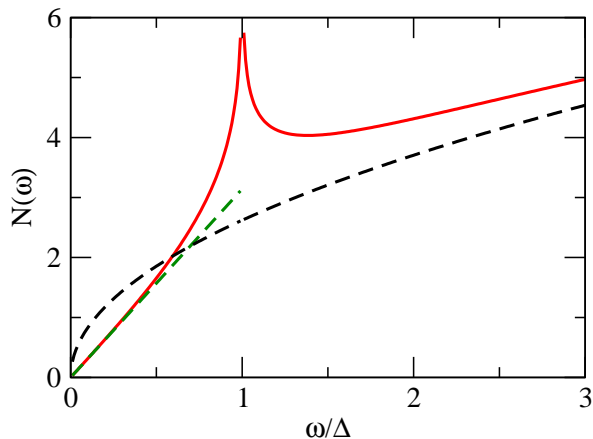}
\caption{The electronic density of states $N(\omega)$ in units of $N_f\sqrt{2m\Delta}/(\pi^2\hbar^2 v)$ based on Eq.~(\ref{eq:dos}) (solid red curve). The dashed green curve is the $\omega/\Delta$ small limit Eq.~(\ref{eq:dosdirac}) which corresponds to ordinary Dirac and the dashed black is for the semi-Dirac model [$\Delta=0$ in the Hamiltonian of Eq.~(\ref{eq:Hmatrix})] and given in Eq.~(\ref{eq:dossemidirac}).
}\label{fig4}
\end{figure}

Results for the density of states $N(\omega)$ in units of $N_f\sqrt{2m\Delta}/(\pi^2\hbar^2 v)$ as a function of $\omega/\Delta$ are presented in Fig.~\ref{fig4}. The solid red curve is based on the numerical evaluation of Eq.~(\ref{eq:dos}). Note the van Hove singularity at $\omega=\Delta$. The dashed green line applies in the limit $\omega/\Delta\to 0$ [Eq.~(\ref{eq:dosdirac})] and the dashed black curve applies in the limit $\omega/\Delta\to \infty$ which is given by Eq.~(\ref{eq:dossemidirac}) and is the result for a semi-Dirac material. We see that the large $\omega$ asymptotic limit has not yet been reached in the full results (red curve). Returning to the conductivity of Fig.~\ref{fig3}, we note that this quantity depends on velocity factors [Eq.~(\ref{eq:v})] not part of the density of states.
Also, it is more closely related to a joint density of states for initial and final states involved in an optical transition, so we expect the van Hove singularity to be at $\Omega=2\Delta$.
 For the $yy$ direction the velocity involved is a constant $v$ while for the $xx$ direction, it involves $\hbar k_x/m$ which varies with $k_x$ and is zero at $k_x=0$. The van Hove singularity comes from the region $k_x\sim 0$ in $k$-space and the factor $\hbar k_x/m$
ensures that this singularity is modified into a simple change in slope
for $\sigma_{xx}^{\rm inter}(\Omega)$. The dashed black curves in Fig.~\ref{fig3} are for comparison with our merging Dirac points model and are for the semi-Dirac limit.\cite{Carbotte:2019} In this model, the dependence on $\Omega$ goes as\cite{Ziegler:2017,Carbotte:2019} $\sigma_{xx}\sim \sqrt{\Omega}$
and $\sigma_{yy}\sim 1/\sqrt{\Omega}$ as seen here.

\begin{figure}
\includegraphics[width=0.9\linewidth]{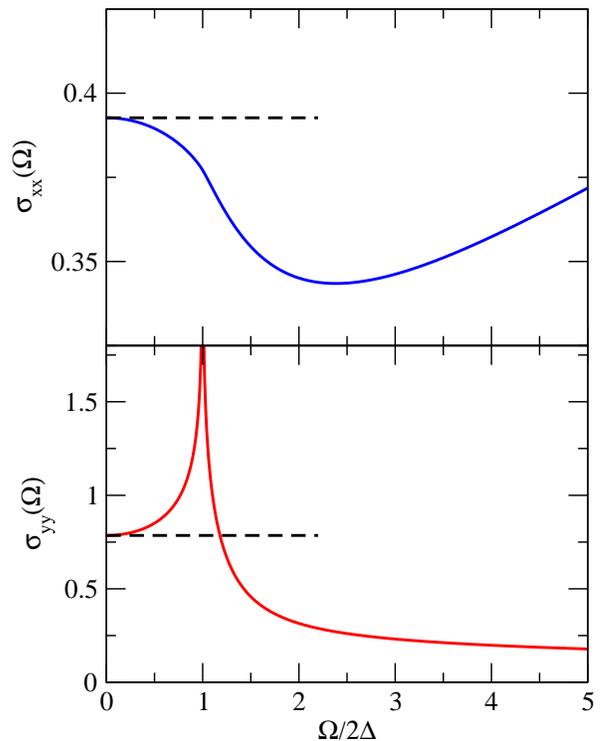}
\caption{The interband optical conductivity $\sigma^{\rm inter}(\Omega)$ as a function of $\Omega/2\Delta$ in a limited range $\Omega/2\Delta \le 5$. The units on $\sigma^{\rm inter}_{xx}(\Omega)$ 
are $N_f(e^2/h)2\sqrt{2\Delta/mv^2}$ (blue curve, top frame) and $N_f(e^2/h)\sqrt{mv^2/2\Delta}$ (red curve, bottom frame) for
  $\sigma^{\rm inter}_{yy}(\Omega)$. The dashed black horizontal lines apply to the pure Dirac limit discussed in the text and agree perfectly with the solid lines at small $\Omega/2\Delta$. 
}\label{fig5}
\end{figure}

In Fig.~\ref{fig5}, we plot again $\sigma^{\rm inter}_{xx}(\Omega)$ (top frame) and $\sigma^{\rm inter}_{yy}(\Omega)$ (bottom frame) but now on a reduced range for $\Omega/2\Delta$ extending only to 5. This emphasizes the small $\Omega$ region where ${\cal F}_{xx}(\bar\Omega)=\pi/8$ and ${\cal F}_{yy}(\bar\Omega)=\pi/4$ [Eqs.(\ref{eq:Fxx}) and (\ref{eq:Fyy}), black dashed lines]. While for $\Omega\to 0$, results for the Dirac limit hold, this does not extend very far. By $\Omega/2\Delta=0.5$, the deviations are already of order 10\%. These deviations capture the effect of a variable $\Delta$ as the merger of the two Dirac points proceeds and is central to our model Hamiltonian in Eq.~(\ref{eq:Hmatrix}).

\begin{figure}
\includegraphics[width=0.9\linewidth]{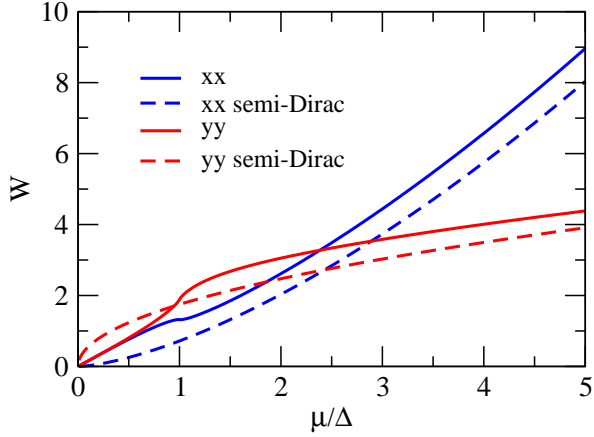}
\caption{The intraband optical conductivity spectral weight $W$ as a function of $\mu/\Delta$. The units are $2\Delta N_f (e^2/h)\sqrt{2\Delta/mv^2}$ for $xx$ and $N_f(e^2/h)\sqrt{2mv^2\Delta}$ for $yy$. The corresponding dashed curves are for the semi-Dirac case when $\Delta=0$ in the Hamiltonian. The units on the square root quantity are $N_f2\Delta e^2/h$. In these units, $W_{xx}$ (dashed blue) is $C_+^{xx}\bar\mu^{3/2}/2$, $W_{yy}$ (dashed red) is $C^{yy}_-\sqrt{\bar\mu}/2$ and the slope of the solid red and solid blue curves at small $\mu/\Delta$ is $\pi\bar\mu/2$, which merges with the Dirac model result.
}\label{fig6}
\end{figure}

Next we consider the intraband (Drude) contribution to the conductivity given by Eq.~(\ref{eq:sxxintra}) and (\ref{eq:syyintra}) for the $xx$ and $yy$ directions, respectively. The Drude weight $W$ of the delta function is shown in Fig.~\ref{fig6} as blue and red solid curves as a function of normalized chemical potential $\mu/\Delta$ at $T=0$ with the units set by the prefactors outside the curly brackets in Eqs.~(\ref{eq:sxxintra}) and (\ref{eq:syyintra}). In the conductivity there is a Dirac delta function, while for the optical spectral weight, the delta function is replaced by a factor of 1/2. Structures in these curves are expected at $\mu/\Delta=1$ analogous to what we saw in the interband optical transitions (in which case the structures are at $\Omega/2\Delta=1$). The red curve shows a sudden increase while the blue curve shows a more modest decrease. Also shown for comparison, as dashed curves, are the asymptotic results given in Eqs.~(\ref{eq:Gxx}) and (\ref{eq:Gyy}) for the ${\cal G}_{xx}$ and ${\cal G}_{yy}$ functions in the $\bar\mu>>1$ limit. These forms correspond to the semi-Dirac limit which results when $\Delta$ is taken to be zero. We have plotted these over the entire range of $\mu/\Delta$ from 0 to 5. While we expect that at larger $\bar\mu$ these forms will merge with our complete numerical results (solid curve) this has not yet occurred at $\bar\mu=5$ where the asymptotic forms are lower in magnitude by order of 10\%, with variation, however, close to the numerical curves. In the energy range around $\mu/\Delta=1$ to 2, the red and blue curves have distinctive behaviors which relate directly to the merging of the two Dirac points into one and also provide a measure of the gap. As the gap is reduced, the doping level needed to sample this critical region is reduced and the intraband transitions could  be used to probe the phase transition which occurs at $\Delta=0$. Note that below $\sim\bar\mu=1$, the semi-Dirac curves deviate very strongly from the complete numerical curves which merge and become linear over a significant region below $\bar\mu=1$. Instead, in semi-Dirac 
$\sigma^{\rm intra}_{xx}$ goes like $(\mu/\Delta)^{3/2}$ and
$\sigma^{\rm intra}_{yy}$ like $(\mu/\Delta)^{1/2}$ (the ratio of which goes as $\mu$ as was also previously shown\cite{Banerjee:2012,Carbotte:2019}). 

\begin{figure}
\includegraphics[width=0.9\linewidth]{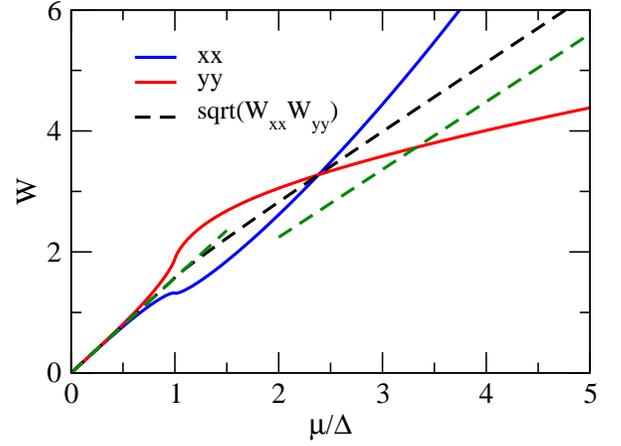}
\caption{The intraband optical conductivity spectral weight $W$ as a function of $\mu/\Delta$ with units as described in Fig.~\ref{fig6}. The dashed black curve  gives the spectral weight for the quantity $\sqrt{W_{xx}W_{yy}}$ in units of $N_f2\Delta e^2/h$, and is obtained from the product of the solid red and solid blue curves. The dashed green lines are the pure Dirac case of $\pi\bar\mu/2$ at small $\mu/\Delta$ and the semi-Dirac limit of $\sqrt{C_+^{xx}C_-^{yy}}\bar\mu/2$ at larger values of $\mu/\Delta\ge 2$.
}\label{fig7}
\end{figure}

In Fig.~\ref{fig7}, we present results  for the square root quantity 
$\sqrt{W_{xx}W_{yy}}$ 
in units of $N_f2\Delta e^2/h$ (dashed black curve) which we compare with 
$W_{xx}$ (blue curve) and $W_{yy}$ (red curve). This new curve (dashed black) is linear at small $\mu/\Delta$ and identical to the red and blue results. All three curves have slope $\pi/2$. Beyond $\mu/\Delta\sim 2$, the dashed black curve again becomes nearly linear with slope close to its semi-Dirac value of 
$\sqrt{C_+^{xx}C_-^{yy}}/2$ 
(dashed green curve).

\begin{figure}
\includegraphics[width=0.9\linewidth]{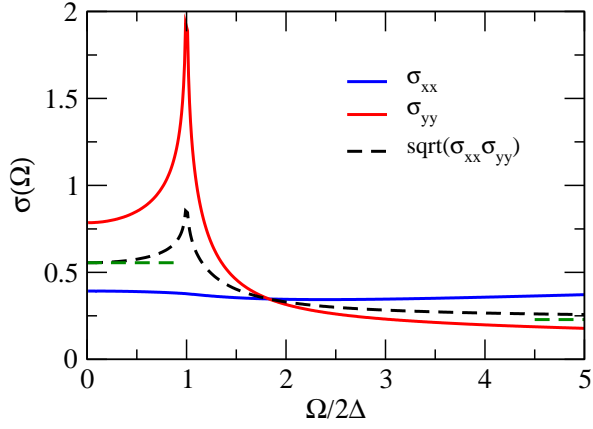}
\caption{The interband contribution to the dynamic conductivity as a function of photon energy $\Omega/2\Delta$ in the same units as Fig.~\ref{fig3}. Here only the range up to $\Omega/2\Delta=5$ is shown and we compare $\sigma_{xx}$ 
(blue curve) and $\sigma_{yy}$ (red curve) with the quantity $\sqrt{\sigma_{xx}\sigma_{yy}}$, this latter having units of $N_f\sqrt{2}e^2/h$ (dashed black curve). The dashed green curves at small photon energy and near $\Omega/2\Delta\lesssim 5$ are for the known Dirac and semi-Dirac limits\cite{Carbotte:2019} equal to $\pi/(4\sqrt{2})$ at small $\Omega/2\Delta$ and $\sqrt{C_-^{xx}C_+^{yy}}/(4\sqrt{2})$ around $\Omega/2\Delta\lesssim 5$, respectively.
}\label{fig8}
\end{figure}

It is of interest to compare the results of Fig.~\ref{fig7} for
$\sqrt{W_{xx}W_{yy}}$  with similar results for the interband transitions. In Fig.~\ref{fig8}, we show
$\sqrt{\sigma_{xx}^{\rm inter}\sigma_{yy}^{\rm inter}}$  versus $\Omega/2\Delta$ 
in units of $\sqrt{2}e^2/h$ (black dashed curve) with our previous results for 
$\sigma_{xx}^{\rm inter}(\Omega)$ 
in units of $(e^2/h)2\sqrt{2\Delta/mv^2}$ (solid blue) and $\sigma_{yy}^{\rm inter}(\Omega)$ in units of $(e^2/h)\sqrt{mv^2/2\Delta}$ (solid red). It is clear that the prominent van Hove singularity in this last curve at $\Omega/2\Delta=1$ is reflected in the results for the square root of the product (dashed black).
This is in contrast to the dashed black curve of Fig.~\ref{fig7} for the chemical potential ($\mu$) variation of the Drude spectral weight. This latter quantity, in comparison, shows little structure at $\mu/\Delta\approx 1$.
 Note also that on the scale chosen for Fig.~\ref{fig8}, the conductivity 
$\sigma_{xx}^{\rm inter}(\Omega)$  shows little variation with photon energy. The dashed green horizontal line at small $\Omega$ is for comparison and gives the Dirac limit equal to $\pi/(4\sqrt{2})$ which agrees with the dashed black curve, while for large $\Omega/2\Delta$, the dashed green horizontal line is for semi-Dirac at
$\sqrt{C_+^{xx}C_-^{yy}}/(4\sqrt{2})$ 
and this also agrees well with the dashed black curve. 
We see that 
$\sqrt{\sigma_{xx}^{\rm inter}\sigma_{yy}^{\rm inter}}$  versus $\Omega/2\Delta$ 
has evolved from Dirac to semi-Dirac but the results around $\Omega/2\Delta=1$ deviate substantially from these two limiting cases and reflect the effect of a finite value of $\Delta$ on the conductivity
$\sqrt{\sigma_{xx}^{\rm inter}\sigma_{yy}^{\rm inter}}$.
Note results for the ratio
$\sigma_{xx}^{\rm inter}(\Omega)/\sigma_{yy}^{\rm inter}(\Omega)$ 
in the limit of 
$\Omega/2\Delta\to 0$ 
and 
$\Omega/2\Delta\to \infty$ 
are, respectively, $2\Delta/mv^2$ and $(C_-^{xx}/C_+^{yy})/(\Omega/mv^2)$.
For $\sigma_{xx}^{\rm inter}(\mu)/\sigma_{yy}^{\rm inter}(\mu)$ in the limit of
$\mu/\Delta\to 0$ and 
$\mu/\Delta\to \infty$, it is
$2\Delta/mv^2$ and $(2\mu/mv^2)(C_+^{xx}/C_-^{yy})$, respectively. 
Of particular importance is the fact that  
$\sigma_{xx}^{\rm inter}(\Omega)/\sigma_{yy}^{\rm inter}(\Omega)$ 
at $\Omega=0$ is equal to $2\Delta/mv^2$ which is a simple ratio of the two energy scales in our Hamiltonian, namely, $2\Delta$ and $mv^2$. For fixed value of $mv^2$, small $\Delta$ implies that
$\sigma_{xx}^{\rm inter}(\Omega=0)$ is much smaller than is
$\sigma_{yy}^{\rm inter}(\Omega=0)$, a fact which can be used to monitor the approach to the phase transition at $\Delta=0$. At large values of $\Omega/2\Delta$, 
$\sigma_{xx}^{\rm inter}(\Omega)/\sigma_{yy}^{\rm inter}(\Omega)$  reduces to $(C_-^{xx}/C_+^{yy})(\Omega/mv^2)$ which could provide a measure of the scale $mv^2$.

\section{Transport: dc conductivity and Wiedemann-Franz}

In the clean limit, the dc conductivity as a function of temperature is given by
Eq.~(\ref{eq:sigmaxxintra}) and (\ref{eq:sigmayyintra}) with the Dirac delta function $\delta(\Omega)$ replaced by $1/2\Gamma$, where $\Gamma$ is the scattering rate. After simplifications, $\sigma^{\rm dc}(T)$ can be written in terms of the universal functions ${\cal G}^{xx}(\bar\mu)$ and ${\cal G}^{yy}(\bar\mu)$, defined the expressions given inside the curly brackets of Eq.~(\ref{eq:sxxintra}) and (\ref{eq:syyintra}), respectively. We obtain
\begin{equation}
\displaystyle
\sigma^{\rm dc}_{xx}(T)=N_f{e^2\over h}{1\over \Gamma \bar T}{\Delta^{3/2}\over\sqrt{2mv^2}}\int_0^\infty{d\bar\omega {\cal G}_{xx}(\bar\omega)\over\cosh^2(\bar\omega/2\bar T)},
\label{eq:sxxdc}
\end{equation}
where we have assumed charge neutrality (no doping or $\mu=0$). Here, $\bar\mu=\mu/\Delta$, $\bar T= T/\Delta$, and $\bar\omega=\omega/\Delta$. Likewise,
\begin{equation}
\displaystyle
\sigma^{\rm dc}_{yy}(T)=N_f{e^2\over h}{1\over 4\Gamma \bar T}\sqrt{2mv^2\Delta}\int_0^\infty{d\bar\omega {\cal G}_{yy}(\bar\omega)\over\cosh^2(\bar\omega/2\bar T)}.
\label{eq:syydc}
\end{equation}
The thermal conductivity $\kappa(T)$ as a function of temperature $T$ or more precisely $\kappa_{xx}(T)/T$ and $\kappa_{yy}(T)/T$ follows from
Eqs.~(\ref{eq:sxxdc}) and (\ref{eq:syydc}) with an additional factor of $(\bar\omega/\bar T)^2$ included in the integration over $\bar\omega$ and the factor of electric charge squared is dropped. The Lorenz number $L(T)$ is related in the usual way to $\sigma^{\rm dc}(T)$ and $\kappa(T)/T$ and is given by
\begin{equation}
L(T)=\displaystyle {\kappa(T)\over T\sigma^{\rm dc}(T)}.
\label{eq:L}
\end{equation}
We note that in this work $L(T)$ will depend on direction ($x,x$) or ($y,y$) as does the electrical conductivity $\sigma^{\rm dc}(T)$ and the thermal conductivity $\kappa(T)$. For the numerical evaluation of Eqs.~(\ref{eq:sxxdc}) to (\ref{eq:L}), it is convenient to introduce new functions $G_1^{xx}(\bar T)$ and $G_2^{xx}(\bar T)$ and equivalent quantities for the $yy$ direction with
\begin{equation}
G_1(\bar T)=\int_0^\infty\displaystyle{dx\over\cosh^2x}{\cal G}(\bar\omega=2\bar Tx)
\label{eq:G1}
\end{equation}
and
\begin{equation}
G_2(\bar T)=\int_0^\infty\displaystyle{x^2dx\over\cosh^2x}{\cal G}(\bar\omega=2\bar Tx),
\label{eq:G2}
\end{equation}
where
for convenience we have suppressed the directional index $xx$ and $yy$. The function $G_1(\bar T)$ enters the dc electrical conductivity while $G_2(\bar T)$ enters the thermal conductivity. Specifically, $\sigma^{\rm dc}(T)$ of Eqs.~(\ref{eq:sxxdc}) and (\ref{eq:syydc}) take the form
\begin{eqnarray}\displaystyle
\sigma_{xx}^{\rm dc}(T)&= N_f \displaystyle{e^2\over h} {2\over\Gamma}{\Delta^{3/2}\over\sqrt{2mv^2}}G_1^{xx}(\bar T),\label{eq:Gcondx}
\\
\sigma_{yy}^{\rm dc}(T)&= N_f \displaystyle{e^2\over h} {1\over2\Gamma}\sqrt{2mv^2\Delta}G_1^{yy}(\bar T)\label{eq:Gcondy}
\end{eqnarray}
and
\begin{eqnarray}\displaystyle
{\kappa_{xx}(T)\over T}&= \displaystyle
N_f {1\over h} {8\over\Gamma}{\Delta^{3/2}\over\sqrt{2mv^2}}
G_2^{xx}(\bar T),\label{eq:Gkappax}\\
{\kappa_{yy}(T)\over T}&= \displaystyle
N_f {1\over h} {2\over\Gamma}\sqrt{2mv^2\Delta}G_2^{yy}(\bar T).\label{eq:Gkappay}
\end{eqnarray}
The Lorenz numbers in the $xx$ and $yy$ directions are then
\begin{eqnarray}\displaystyle
{L_{xx}(T)}&= \displaystyle{4\over e^2} {G_2^{xx}(\bar T)\over G_1^{xx}(\bar T)},
\label{eq:GLx}\\
{L_{yy}(T)}&= \displaystyle{4\over e^2} {G_2^{yy}(\bar T)\over G_1^{yy}(\bar T)}.
\label{eq:GLy}
\end{eqnarray}

\begin{figure}
\includegraphics[width=0.9\linewidth]{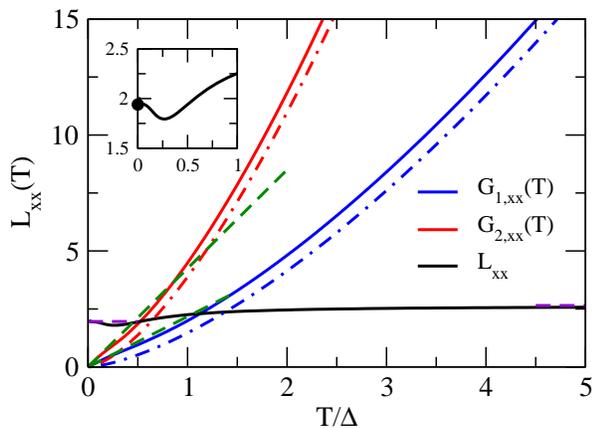}
\caption{The $G$ functions related directly to the temperature dependent dc conductivity $\sigma^{\rm dc}(T)$ by Eqs.~(\ref{eq:Gcondx}) and (\ref{eq:Gcondy}), 
to the thermal conductivity $\kappa(T)$ through
Eqs.~(\ref{eq:Gkappax}) and (\ref{eq:Gkappay}),
and to the Lorenz number $L(T)$ by
Eqs.~(\ref{eq:GLx}) and (\ref{eq:GLy}). The solid blue curve is for   
for the dc conductivity $\sigma_{xx}^{\rm dc}(T)$ and solid red is for the thermal conductivity $\kappa_{xx}(T)/T$. The corresponding  dash-dotted curves are for comparison to the results of the complete calculation and are for the semi-Dirac case. The dashed green curves are the pure Dirac case, both are linear in $T/\Delta$ and, for a limited range around $T/\Delta\sim 0$, agree well with the full calculations for our merging Dirac point model. The dashed purple lines around $T/\Delta\sim 0$ and around $T/\Delta\sim 5$ agree well with the solid black curve and are the results for the Lorenz number in the Dirac and semi-Dirac cases, respectively. The inset is a blow up of the small $T/\Delta$ region for the Lorenz number.
}\label{fig9}
\end{figure}

\begin{figure}
\includegraphics[width=0.9\linewidth]{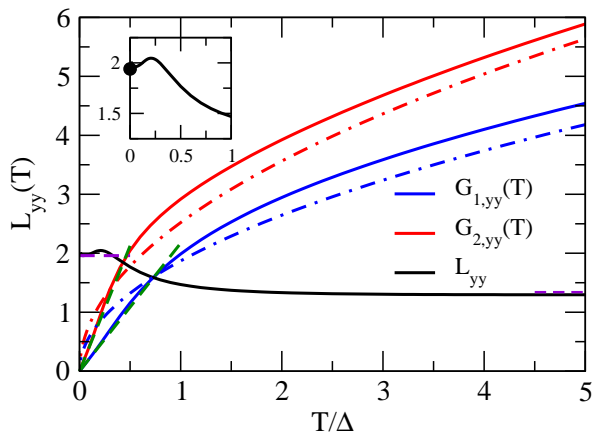}
\caption{Same as in Fig.~\ref{fig9} but now the results are for the $yy$ direction.
}\label{fig10}
\end{figure}

In Fig.~\ref{fig9} and \ref{fig10}, we present our numerical results for the transport in the $x$ and $y$ directions, respectively.
What is plotted are the functions $G_1$, $G_2$ and $G_2/G_1$. The curves can be interpreted directly in terms of electrical ($G_1$) and thermal ($G_2$) conductivity and Lorenz number. In this case the units on $\sigma_{xx}^{\rm dc}$ are
$N_f {e^2\over h} {2\over\Gamma}{\Delta^{3/2}\over\sqrt{2mv^2}}$ and on 
$\sigma_{yy}^{\rm dc}$ are
$N_f {e^2\over h} {1\over 2\Gamma}\sqrt{2mv^2\Delta}$, while for
$\kappa(T)/T$ they are, respectively, 
$N_f {1\over h} {8\over\Gamma}{\Delta^{3/2}\over\sqrt{2mv^2}}$
and
$N_f {1\over h} {2\over\Gamma}\sqrt{2mv^2\Delta}$. For $L$ the units are $4/e^2$. The solid red curves are for $G_2(\bar T)$ and solid blue for $G_1(\bar T)$, while the solid black gives $L(T)$. The $xx$ and $yy$ results are given in Figs.~\ref{fig9} and \ref{fig10}, respectively.

Analytic results can be obtained in the low temperature limit. In this case, 
the integrands in  Eqs.~(\ref{eq:G1}) and (\ref{eq:G2}) are highly peaked about $\bar\omega\sim 0$ and ${\cal G}(\bar\omega)\simeq\pi\bar\omega/2$ [see Eq.~(\ref{eq:Gxx}) and (\ref{eq:Gyy})]. We obtain
\begin{eqnarray}
G_1(\bar T)&=\pi\bar T\int_0^\infty\displaystyle{x dx\over\cosh^2x}=\pi\bar T\ln 2=2.18\bar T,
\label{eq:G1no}\\
G_2(\bar T)&=\pi\bar T\int_0^\infty\displaystyle{x^3dx\over\cosh^2x}=1.35 \pi\bar T = 4.24\bar T.
\label{eq:G2no}
\end{eqnarray}
The dashed green lines in both Figs.~\ref{fig9} and \ref{fig10} show that this agrees with our numerical work but that the asymptotic forms hold only over a small temperature interval. It is also interesting to compare our results with those based on the Hamiltonian in Eq.~(\ref{eq:Hmatrix}) but with $\Delta=0$ which corresponds to the semi-Dirac limit. In this case, 
${\cal G}_{xx}(\bar\omega)=C_+^{xx}\bar\omega^{3/2}/2$
and
${\cal G}_{yy}(\bar\omega)=C_-^{yy}\sqrt{\bar\omega}/2$
over the entire energy range while it also applies to the $\bar\omega\to\infty$ limit of our more complicated model as recorded in Eq.~(\ref{eq:Gxx}) and
(\ref{eq:Gyy}). We obtain
\begin{eqnarray}
G_1^{xx}(\bar T)&=C_+^{xx}\sqrt{2}\bar T^{3/2}\int_0^\infty\displaystyle{x^{3/2}dx\over\cosh^2x}=1.46\bar T^{3/2},\\
G_2^{xx}(\bar T)&=C_+^{xx}\sqrt{2}\bar T^{3/2}\int_0^\infty\displaystyle{x^{7/2}dx\over\cosh^2x}=3.88\bar T^{3/2},\\
G_1^{yy}(\bar T)&=\displaystyle{C_-^{yy}\over\sqrt{2}}\bar T^{1/2}\int_0^\infty\displaystyle{x^{1/2}dx\over\cosh^2x}=1.87\bar T^{1/2},\\
G_2^{yy}(\bar T)&=\displaystyle{C_-^{yy}\over\sqrt{2}}\bar T^{1/2}\int_0^\infty\displaystyle{x^{5/2}dx\over\cosh^2x}=2.52\bar T^{1/2},
\end{eqnarray}
for the semi-Dirac model. These results are shown in Figs.~\ref{fig9} and \ref{fig10}
for comparison with our numerical results based on our Hamiltonian with finite $\Delta$. They are given as the dash dotted red curve for the thermal conductivity and the blue for the electrical conductivity. We see that solid and dash-dotted curves track each other reasonably well over the entire temperature range shown but there are differences. The $T/\Delta\to 0$ limits are different and the magnitude of the curves also change. This shows that finite $\Delta$ results differ somewhat from semi-Dirac.

Finally, we turn to the solid black curves in Fig.~\ref{fig9} and \ref{fig10} which relate to the Lorenz number given by the ratio $G_2(\bar T)/G_1(\bar T)$ in our chosen units of $4/e^2$. In the Dirac limit which applies to our model at very low temperature, we get 1.94 shown as the purple dashed curve and this applies to both directions. For the semi-Dirac limit, we get 2.66 for $xx$ and 1.35 for $yy$, again shown as purple dashed curves around $\bar T=5$. This holds for any $\bar T$ in semi-Dirac. We note that our numerical results (black curve) agree well with the limit at the higher temperatures shown but at intermediate temperatures there are differences. In particular, the differences are quite significant in the low temperature region shown in the inset to each of Fig.~\ref{fig9} and \ref{fig10}. Since we are in the clean limit, the temperature variations shown in the Lorenz number are due entirely to bandstructure effects. Over the entire temperature range shown $L_{xx}$ starts at 1.94 near $\bar T=0$ and rises to 2.65 at $\bar T=5$ (an increase), while $L_{yy}$ goes from 1.94 to 1.35 (a decrease). In a recent preprint, Lavasani {\it et al.}\cite{Lavasani:2019} have emphasized that the electron-phonon interaction can lead to violations of the Wiedemann Franz law even in a Fermi liquid. Here we find such violation due entirely to bandstructure effects.

\section{Discussions and conclusions}

We have examined a model Hamiltonian for merging Dirac points which is characterized by a velocity $v$, mass $m$ and energy gap $\Delta$.
We have shown that the interband optical conductivity in the $xx$ direction at zero photon energy is equal to $(\pi e^2/4h)\sqrt{2\Delta/mv^2}$ and in the $yy$ direction $(\pi e^2/4h)\sqrt{mv^2/2\Delta}$ 
so that the square root of their product $\sqrt{\sigma_{xx}^{\rm inter}(0)\sigma_{yy}^{\rm inter}(0)}=\pi e^2/4h$, 
which is exactly the value of the graphene universal constant value accounting for degeneracy factors. This is the relativistic Dirac limit. The dimensionless factors
 $\sqrt{mv^2/2\Delta}$ which cancel out of the product quantity show that the ratio 
$\sigma_{xx}^{\rm inter}(0)/\sigma_{xx}^{\rm inter}(0)$ goes like $2\Delta/mv^2$ and changes from a value smaller than one as $2\Delta< mv^2$ to a value larger than one for $2\Delta>mv^2$. Consequently, this ratio could be used to obtain information on the relative size of the gap as compared with $mv^2$.

As the photon energy $\Omega$ is increased out of zero, the interband background changes in contrast to graphene where it remains constant. For large $\Omega/2\Delta$ it evolves to the semi-Dirac value which is $(e^2/4h)C_-^{xx}\sqrt{\Omega/mv^2}$ for $xx$
and $(e^2/4h)C_+^{yy}\sqrt{mv^2/\Omega}$ for $yy$. That is, proportional to the square root of $\Omega$ for $xx$ and inversely for $yy$. Here, $C_-^{xx}=0.902$ and $C_+^{yy}=1.75$. Note that in this limit, the ratio 
$\sigma_{xx}^{\rm inter}(\Omega\to\infty)/\sigma_{yy}^{\rm inter}(\Omega\to\infty)=(C_-^{xx}/C_+^{yy})(\Omega/mv^2)$ which is proportional to photon energy normalized to the characteristic energy $mv^2$ associated with our Hamiltonian. There is no dependence on the energy gap $\Delta$ in this limit.

Between the small (Dirac) and the large (semi-Dirac-like) $\Omega$ limits, there are large deviations from both these limiting behaviors which provide information on how the phase transition from finite $\Delta$ to $\Delta=0$ manifests in the conductivity. The conductivity 
$\sigma^{\rm inter}_{yy}(\Omega)$
displays a large van Hove singularity at $\Omega=2\Delta$ while
$\sigma^{\rm inter}_{xx}(\Omega)$ only shows a change in slope from concave downward to upward. The square root of the product of these two quantities also shows a van Hove singularity at $\Omega=2\Delta$. At small $\Omega$, it agrees with the pure relativistic Dirac limit and at large, with semi-Dirac\cite{Carbotte:2019}.

For the intraband contribution to the conductivity (Drude component), we find it to have optical spectral weight at chemical potential $\mu\to 0$ of 
$(\pi e^2/h)\sqrt{2\Delta/mv^2}\mu$  and $(\pi e^2/h)\sqrt{mv^2/2\Delta}\mu$  for
$\sigma_{xx}^{\rm intra}$ and $\sigma_{yy}^{\rm intra}$.
The material factor $\sqrt{2\Delta/mv^2}$ drops out of the square root of their product which behaves as in graphene and is equal to $e^2\pi\mu/h$. As $\mu$ is increased, the intraband conductivity evolves into its value for semi-Dirac\cite{Carbotte:2019} (a $\mu^{3/2}$ behavior).
Between the small and large $\mu$ regime, large deviations are predicted which carry the information on the evolution of the system between these two limits as seen in optical experiments. But these are much smaller in the Drude weight as a function of $\mu$ than they are in the longitudinal dynamic conductivity as a function of photon energy $\Omega/2\Delta$.

Transport properties have also been studied as a function of temperature $T$. The dc conductivity is linear in $T$ at low temperature with the material parameter $\sqrt{2\Delta/mv^2}$ dropping out of the square root of the product 
$\sqrt{\sigma_{xx}^{\rm intra}(T\to 0)\sigma_{yy}^{\rm intra}(T\to 0)}=2\pi T\ln 2 (e^2/h)$, which is the value for a pure 2-dimensional relativistic material. As temperature is increased, the dc conductivity evolves towards the $T^{3/2}$ (for $xx$) and $T^{1/2}$ (for $yy$) behavior of  semi-Dirac with $\Delta=0$, but remains somewhat below the full numerical calculations, a somewhat different magnitude in the temperature range considered up to $T/\Delta=5$.  The electronic thermal conductivity evolves similarly. The Lorenz number shows temperature dependence. $L^{xx}$ starts at the graphene value of $2.4L_0$, with $L_0=\pi^2/3e^2$, then has a dip around $T/\Delta\sim 0.25$ and then increases gradually towards the semi-Dirac value for this direction which is $3.3L_0$. $L^{yy}$ starts at the same value of $2.4L_0$ as $T\to 0$ but has a maximum around $T/\Delta\sim 0.25$ and then slowly decreases towards the value $1.67L_0$, the semi-Dirac value for the $yy$ direction, which is half the $xx$ value. The square root of the product $\sqrt{L^{xx}L^{yy}}\simeq 2.4L_0$, which is the same as for graphene and this quantity is nearly independent of temperature.

In conclusion, we have elucidated a variety of the features of merging Dirac points in optics and transport  which may aid in identifying such a system and deducing key model parameters.

\begin{acknowledgments}
We thank Kyle Bryenton for assistance in producing Fig.~\ref{fig1}.
This work has been supported by the Natural Sciences and Engineering Council of Canada (NSERC) and by the
Canadian Institute for Advanced Research (CIFAR).

\end{acknowledgments}


%

\end{document}